\documentclass[twocolumn,aps,superscriptaddress,showpacs,nofootinbib,floatfix]{revtex4}
\usepackage{epsfig,bm,feynmf}
\usepackage{graphics}
\usepackage{amsmath}
\usepackage[normalem]{ulem}  
\usepackage[dvips]{color} 

\renewcommand\sout{\bgroup \color{red} \ULdepth=-.5ex \ULset}

\begin{document}


\title{Effects of medium modification of pion production threshold in heavy ion collisions and the nuclear symmetry energy}


\author{Taesoo Song}\email{song@fias.uni-frankfurt.de}
\affiliation{Frankfurt Institut for Advanced Studies and Institute for Theoretical Physics, Johann Wolfgang Goethe Universit¨at, Frankfurt am Main, Germany}
\author{Che Ming Ko}\email{ko@comp.tamu.edu}
\affiliation{Cyclotron Institute and Department of Physics and Astronomy, Texas A$\&$M University, College Station, TX 77843-3366, USA}


\begin{abstract}
Using the relativistic Vlasov-Uehling-Uhlenbeck (RVUU) equation based on mean fields from the nonlinear relativistic models, we study the effect of medium modification of pion production threshold on the total pion yield and the $\pi^-/\pi^+$ ratio in Au+Au collisions. We find that the in-medium threshold effect enhances both the total pion yield and the $\pi^-/\pi^+$ ratio, compared to those without this effect. Furthermore, including the medium modification of the pion production threshold in asymmetric nuclear matter leads to a larger $\pi^-/\pi^+$ ratio for the $NL\rho\delta$ model with a stiffer symmetry energy than the $NL\rho$ model with a softer symmetry energy, opposite to that found without the in-medium threshold effect. Experimental data from the FOPI Collaboration are reproduced after including a density-dependent cross section for $\Delta$ baryon production from nucleon-nucleon collisions, which suppresses the total pion yield but hardly changes the $\pi^-/\pi^+$ ratio. The large errors in the experimentally measured $\pi^-/\pi^+$ ratio prevent, however, the distinction between the predictions from the $NL\rho$ and $NL\rho\delta$ models.
\end{abstract}


\maketitle

\section{introduction}

The nuclear symmetry energy is an important quantity for describing the properties of asymmetric nuclear matter. It is defined as the second derivative of the energy per nucleon in asymmetric nuclear matter with respect to its isospin asymmetry $\alpha=(\rho_n-\rho_p)/(\rho_n+\rho_p)$, where $\rho_n$ and $\rho_p$ are the neutron and proton densities, respectively. Although our knowledge on the nuclear symmetry energy at normal and subnormal densities have been relatively well determined from the properties of nuclei and isospin sensitive observables in intermediate-energy heavy ion collisions, very little is known about its behaviors at high densities~\cite{Baran:2004ih,Li:2008gp}. In Ref.~\cite{Li:2002qx}, it was suggested that the $\pi^-/\pi^+$ ratio in heavy ion collisions induced by neutron-rich nuclei at energies near the pion production threshold in nucleon-nucleon collisions in free space would be sensitive to the stiffness of nuclear symmetry energy at high densities. This can be understood by noting that pions are produced in these collisions from the decay of Delta resonances created from nucleon-nucleon collisions during the compression stage of heavy ion collisions, with $\Delta^-$ mainly from $n+n$ collisions and $\Delta^{++}$ mainly from $p+p$ collisions~\cite{Huber:1994ee}. Since the stiffness of nuclear symmetry energy affects the isospin asymmetry of produced dense nuclear matter, with a soft one leading to more neutrons than protons, more $\Delta^-$ and thus more $\pi^-$ are produced, resulting in a larger $\pi^-/\pi^+$ ratio in the final pion yield~\cite{Gaitanos:2003zg}. Many theoretical studies based on various transport models have since been carried out to study the $\pi^-/\pi^+$ ratio in heavy ion collisions. Comparing these theoretical results with experimental data from the FOPI Collaboration~\cite{Reisdorf:2006ie} has led, however, to widely different conclusions on the stiffness of the nuclear symmetry energy at high densities. Some of these studies indicate that the nuclear symmetry energy is stiff and increases with nuclear density~\cite{Feng:2009am}, while others claim that it is supersoft and vanishes at about three times the normal nuclear matter density~\cite{Xiao:2009zza,Xie:2013np}.

In all above studies, the cross section for Delta resonance production in nucleon-nucleon collisions is taken from that in free space. As pointed out in Ref.~\cite{Ferrini:2005jw}, the threshold for this reaction is modified in neutron-rich matter when one take into account the effect of nuclear symmetry energy on the mean-field potentials acting on initial and final particles. This effect also changes the branching ratio of a Delta resonance decaying to different pion charged states. As a result, the final $\pi^-/\pi^+$ ratio in a heavy ion collision including such threshold effect is different from that without this effect. In the study of Ref.~\cite{Ferrini:2005jw} based on the relativist mean-field models, it was found that the threshold effect tends to cancel the effect of the symmetry energy on the $\pi^-/\pi^+$ ratio in heavy ion collisions, thus reducing the sensitivity of this ratio to the stiffness of nuclear symmetry energy at high densities.

The threshold effect on pion production studied in Ref.~\cite{Ferrini:2005jw} has, however, not included the comparison with experimental data.  Also, The effect was not included in a fully covariant way. In the present study, we improve the treatment of Ref.~\cite{Ferrini:2005jw} and also compare the results with available data to extract the nuclear symmetry energy at high densities. Specifically, we extend the Relativistic Vlasov-Uhling-Uhlenbeck (RVUU) model~\cite{Ko:1987gp,Ko:1988zz,Ko:1996yy} by including explicitly the different isospin states of nucleons, Delta resonances, and pions. As in Ref.~\cite{Ferrini:2005jw}, our study is based on the nonlinear relativistic mean-field models that describe the interaction between nucleons by the exchange of scalar and vector mesons as well as isoscalar and isovector mesons. We find that the threshold effect due to the vector mean fields cancels or even reverses the effect of nuclear symmetry energy on the $\pi^-/\pi^+$ ratio.  On the other hand, the threshold effect due to the scalar mean fields enhances the pion yield regardless of the stiffness of nuclear symmetry energy, assuming that the cross sections for $\Delta$ production have the same form as in vacuum. To reproduce the experimental data on the pion yield requires, however, the introduction of density dependence in the cross sections for $\Delta$ production in nuclear matter.

This paper is organized as follows: In Sec.~\ref{NLRMF}, we briefly review the nonlinear relativistic mean-field model. We then describe in Sec.~\ref{rvuu} the RVUU equation for the time evolutions of the nucleon, $\Delta$, and pion phase-space distribution functions under the influence of the relativistic mean fields as well as the nucleon and $\Delta$ scatterings, and $\Delta$ decays. The covariant threshold effect is explained in Sec.~\ref{covariant}, and the results from Au+Au collisions are given in Sec.~\ref{results}. Finally, a summary is given in Sec.~\ref{summary}. The derivation of symmetry energy and the discussion on threshold energy in the relativistic mean-field model are given in Appendix~\ref{symmetryE} and \ref{thresholdE}, respectively.

\section{nonlinear relativistic mean-field model}\label{NLRMF}

The Lagrangian for the relativistic mean-field model $NL\rho$ or $NL\rho\delta$ is given by~\cite{Liu:2001iz}
\begin{eqnarray}
L=\bar{N}\bigg[\gamma_\mu(i\partial^\mu -g_\omega\omega^\mu-g_\rho\boldsymbol\tau \cdot \boldsymbol\rho^\mu)\nonumber\\
-(m_N-g_\sigma\sigma-g_\delta \boldsymbol\tau \cdot \boldsymbol\delta)\bigg]N\nonumber\\
+\frac{1}{2}(\partial_\mu \sigma \partial^\mu \sigma-m_\sigma^2\sigma^2)-\frac{a}{3}~\sigma^3-\frac{b}{4}~\sigma^4\nonumber\\
-\frac{1}{4}\Omega_{\mu\nu}\Omega^{\mu\nu}+\frac{1}{2}m_\omega^2\omega_\mu\omega^\mu\nonumber\\
+\frac{1}{2}(\partial_\mu \boldsymbol\delta \partial^\mu \boldsymbol\delta-m_\delta^2\boldsymbol\delta\cdot\boldsymbol\delta)\nonumber\\
-\frac{1}{4}\boldsymbol{R}_{\mu\nu}\cdot\boldsymbol{R}^{\mu\nu}+\frac{1}{2}m_\rho^2\boldsymbol\rho_\mu\cdot\boldsymbol\rho^\mu,
\label{lagrangian}
\end{eqnarray}
where $\Omega_{\mu\nu}=\partial_\mu \omega_\nu- \partial_\nu \omega_\mu$ and $\boldsymbol{R}_{\mu\nu}=\partial_\mu \boldsymbol
\rho_\nu- \partial_\nu \boldsymbol\rho_\mu$. In the Lagrangian, $N$,
$\sigma$, $\omega_\mu$, $\boldsymbol\delta$, and $\boldsymbol\rho_\mu$ denote the nucleon, isoscalar-scalar, isoscalar-vector,
isovector-scalar, and isovector-vector fields, respectively, with their corresponding masses $m_N$, $m_\sigma$,
$m_\omega$, $m_\delta$, and $m_\rho$. The couplings of the mesons to nucleons are given by $g_\sigma$, $g_\omega$, $g_\delta$, and $g_\rho$, while the $\sigma$ meson self interaction is described by the strength parameters $a$ and $b$. Values of the parameters in the $NL\rho$ and $NL\rho\delta$ models are given in Table \ref{parameters}~\cite{Liu:2001iz}. We note that the $\boldsymbol\delta$ field is absent in the $NL\rho$ model.

\begin{table}[h]
\centering
\begin{tabular}{c| c c c c c}
\hline
~~~ & &~$NL\rho$& &$NL\rho\delta$~& \\[2pt]
\hline
~$f_i\equiv (g_i/m_i)^2~$~& & & & & \\[2pt]
~$f_\sigma~({\rm fm^2})$~& & &10.33 & & \\[2pt]
~$f_\omega~({\rm fm^2})$~& & &5.42 & & \\[2pt]
~$f_\rho~({\rm fm^2})$~& &~0.95 & &3.15~ & \\[2pt]
~$f_\delta~({\rm fm^2})$~& &~0 & &2.5~ & \\[2pt]
\hline
~$a/g_\sigma^3~({\rm fm^{-1}})$~& & &0.033 & & \\[2pt]
~$b/g_\sigma^4$~ & & &-0.0048 & & \\[2pt]
\hline
\end{tabular}
\caption{Parameters in the $NL\rho$ and $NL\rho\delta$ models with $f_i$ defined by $(g_i/m_i)^2$~\cite{Liu:2001iz}.}
\label{parameters}
\end{table}

In the mean-field approximation, one neglects the derivatives of meson fields and obtains following field equations:
\begin{eqnarray}
\bigg[\gamma_\mu(i\partial^\mu -g_\omega\omega^\mu-g_\rho\tau_3\rho_3^\mu)\nonumber\\
-(m_N-g_\sigma\sigma-g_\delta \tau_3\delta_3)\bigg]N=0,\label{nucleon}\\
m_\sigma^2 \sigma+a\sigma^2+b\sigma^3=g_\sigma\bar{N}N,\nonumber\\
m_\delta^2 \delta_3=g_\delta\bar{N}\gamma^0\tau_3N,\nonumber\\
m_\omega^2\omega^\mu=g_\omega\bar{N}\gamma^\mu N,\nonumber\\
m_\rho^2\rho_3^\mu=g_\rho\bar{N}\tau_3N.\label{meson1}
\end{eqnarray}
Eq.(\ref{lagrangian}) then becomes the Lagrangian for noninteracting nucleons with effective mass $m_i^*$ and mechanical energy-momentum $p_i^{\mu*}$,
\begin{eqnarray}
m_i^*=m_N-g_\sigma\sigma\mp g_\delta\delta_3,\nonumber\\
p_i^{\mu*}=p^\mu -g_\omega\omega^\mu\mp g_\rho\rho_3^\mu,
\label{momentum}
\end{eqnarray}
where $i=p,n$ for the upper and lower signs, respectively.
In terms of the nucleon scalar and vector densities,
\begin{eqnarray}
\phi_i=\int\frac{d^3{\bf p}_i^*}{(2\pi)^3}\frac{E_i^*}{m_i^*}f_i({\bf p}_i^*),\nonumber\\
j_i^\mu=\int\frac{d^3{\bf p}_i^*}{(2\pi)^3}\frac{p_i^{\mu*}}{E_i^*}f_i({\bf p}_i^*),
\label{source}
\end{eqnarray}
with $E_i^*=\sqrt{m_i^{*2}+{\bf p}_i^{*2}}$ and $f_i({\bf p}_i^*)$ being the nucleon distribution function including the spin degeneracy, the meson field equations can then be expresses as
\begin{eqnarray}
m_\sigma^2 \sigma+a\sigma^2+b\sigma^3= g_\sigma(\phi_p +\phi_n),\nonumber\\
m_\delta^2 \delta_3= g_\delta(\phi_p -\phi_n),\nonumber\\
m_\omega^2\omega^\mu= g_\omega(j_p^\mu+j_n^\mu),\nonumber\\
m_\rho^2\rho_3^\mu= g_\rho(j_p^\mu-j_n^\mu).\label{meson2}
\end{eqnarray}

\begin{figure}[h]
\centerline{
\includegraphics[width=9 cm]{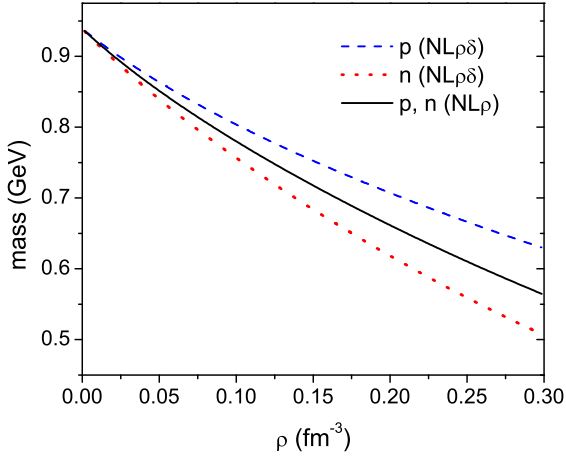}}
\caption{(Color online) Proton and neutron masses in asymmetric nuclear matter of isospin asymmetry $\alpha=0.5$ and at zero temperature as functions of nuclear matter density in the $NL\rho$ and $NL\rho\delta$ models~\cite{Liu:2001iz}.}
\label{mass}
\end{figure}

The nucleon effective mass can be obtained by solving above equations iteratively. Figure \ref{mass} shows the proton and neutron masses in asymmetric nuclear matter of isospin asymmetry $\alpha=(j_n^0-j_p^0)/(j_n^0+j_p^0)=0.5$ and at zero temperature as functions of nuclear matter density for the $NL\rho$ and $NL\rho\delta$ models. It is seen that the proton and neutron masses in asymmetric nuclear matter are degenerate in the $NL\rho$ model but become different in the $NL\rho\delta$ model with the proton mass larger than the neutron mass as a result of their interactions with the isovector-isoscalar meson $\boldsymbol\delta$~\cite{Liu:2001iz}.

In the mean-field approximation, the energy-momentum tensor of a nuclear matter is given by~\cite{Liu:2001iz}
\begin{eqnarray}
T_{\mu\nu}=\frac{\partial L}{\partial(\partial^\mu N)}\partial_\nu N-g_{\mu\nu}L~~~~~~~~~~~~~~~~~~~~~~~~~\nonumber\\
=i\bar{N}\gamma_\mu \partial_\nu N+g_{\mu\nu}\bigg[\frac{1}{2}m_\sigma^2\sigma^2+\frac{a}{3}~\sigma^3+\frac{b}{4}~\sigma^4\nonumber\\
-\frac{1}{2}m_\omega^2\omega_\lambda\omega^\lambda+\frac{1}{2}m_\delta^2\delta_3^2-\frac{1}{2}m_\rho^2\rho_{3\lambda}\rho_3^\lambda\bigg],
\end{eqnarray}
where Eq.~(\ref{nucleon}) has been used in obtaining the expression in the square bracket.

In a static nuclear matter with ${\bf p}_i^*={\bf p}_i$, the energy density and pressure are then
\begin{eqnarray}
\epsilon=T_{00}=\sum_{i=n,p}2\int\frac{d^3{\bf p}_i}{(2\pi)^3}E_i^*f_i({\bf p}_i)\nonumber\\
+\frac{1}{2}m_\sigma^2\sigma^2+\frac{a}{3}~\sigma^3+\frac{b}{4}~\sigma^4+\frac{1}{2}m_\omega^2\omega_0^2\nonumber\\
+\frac{1}{2}m_\delta^2\delta_3^2+\frac{1}{2}m_\rho^2(\rho_{3})_0^2,
\label{energy}
\end{eqnarray}
\begin{eqnarray}
p=\frac{1}{3}T_{jj}=\sum_{i=n,p}\frac{2}{3}\int\frac{d^3{\bf p}_i}{(2\pi)^3}\frac{|{\bf p}_i|^2}{E_i^*}f_i({\bf p}_i)\nonumber\\
+\frac{1}{2}m_\sigma^2\sigma^2+\frac{a}{3}~\sigma^3+\frac{b}{4}~\sigma^4-\frac{1}{2}m_\omega^2\omega_0^2\nonumber\\
+\frac{1}{2}m_\delta^2\delta_3^2-\frac{1}{2}m_\rho^2(\rho_{3})_0^2.
\label{pressure}
\end{eqnarray}

\begin{figure}[h]
\centerline{
\includegraphics[width=9 cm]{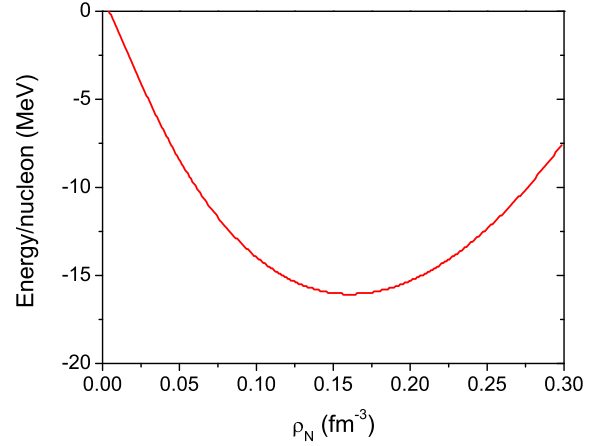}}
\caption{(Color online) Binding energy per nucleon of symmetric nuclear matter at zero temperature as a function of nuclear matter density.}
\label{matter}
\end{figure}

The binding energy of a nucleon in nuclear matter is given by
\begin{eqnarray}
E/N=\epsilon/\rho_N-m_N,
\end{eqnarray}
where $\rho_N=j_n^0+j_p^0$ is the density of the nuclear matter.
For symmetric nuclear matter, both $NL\rho$ and $NL\rho\delta$ models give the same binding energy as a result of vanishing isospin asymmetry. This is shown in Fig.~\ref{matter} for symmetric nuclear matter at zero temperature. It indicates that the binding energy per nucleon of symmetric nuclear matter in both $NL/\rho$ and $NL\rho\delta$ model is 16 MeV at the saturation density of $\rho_0= 0.16 ~{\rm fm^{-3}}$. The curvature of the binding energy at $\rho_0$ is related to the incompressibility of symmetric nuclear matter, given by~\cite{compress}
\begin{eqnarray}
K=9\rho_N^2\frac{\partial^2 (E/N)}{\partial \rho_N^2}\bigg|_{\rho_0}.
\end{eqnarray}
Its value in both $NL\rho$ and $Nl\rho\delta$ models is about 240~MeV.

In asymmetric nuclear matter, the energy per nucleon is a function of the density $\rho_N$ and isospin asymmetry $\alpha$. When expanded in terms of $\alpha$, it is written as
\begin{eqnarray}
E/N(\rho_N,\alpha)=E/N(\rho_N,0)+E_{\rm sym}(\rho_N)\alpha^2+\cdots,
\end{eqnarray}
where
\begin{eqnarray}
E_{\rm sym}(\rho_N)=\frac{1}{2}\frac{\partial^2 (E/N)}{\partial \alpha^2}\bigg|_{\alpha=0}=\frac{1}{2}\rho_N\frac{\partial^2 \epsilon}{\partial \rho_I^2}\bigg|_{\rho_I=0}
\end{eqnarray}
with $\rho_I=j_p^0-j_n^0$ being the isospin density, is the nuclear symmetry energy.

For nuclear matter at zero temperature, the symmetry energy in the $NL\rho$ and $NL\rho\delta$ models are given by~\cite{Kubis:1997ew,Liu:2001iz}
\begin{eqnarray}
E_{\rm sym}(\rho_N)=\frac{p_F^2}{6E_F}+\frac{\rho_N}{2}\bigg[f_\rho-\frac{f_\delta m^{*2}}{E_F^2\{1+f_\delta A(p_F, m^*)\}}\bigg],\nonumber\\
\end{eqnarray}
where $p_F$ and $E_F$ are the nucleon Fermi momentum and energy, respectively, and
\begin{eqnarray}
A(p_F, m^*)=\frac{4}{(2\pi)^3}\int^{p_F}d^3p\frac{p^2}{(p^2+m^{*2})^{3/2}}.
\label{aa}
\end{eqnarray}
Details of the derivation are given in Appendix~\ref{symmetryE}.

\begin{figure}[h]
\centerline{
\includegraphics[width=9 cm]{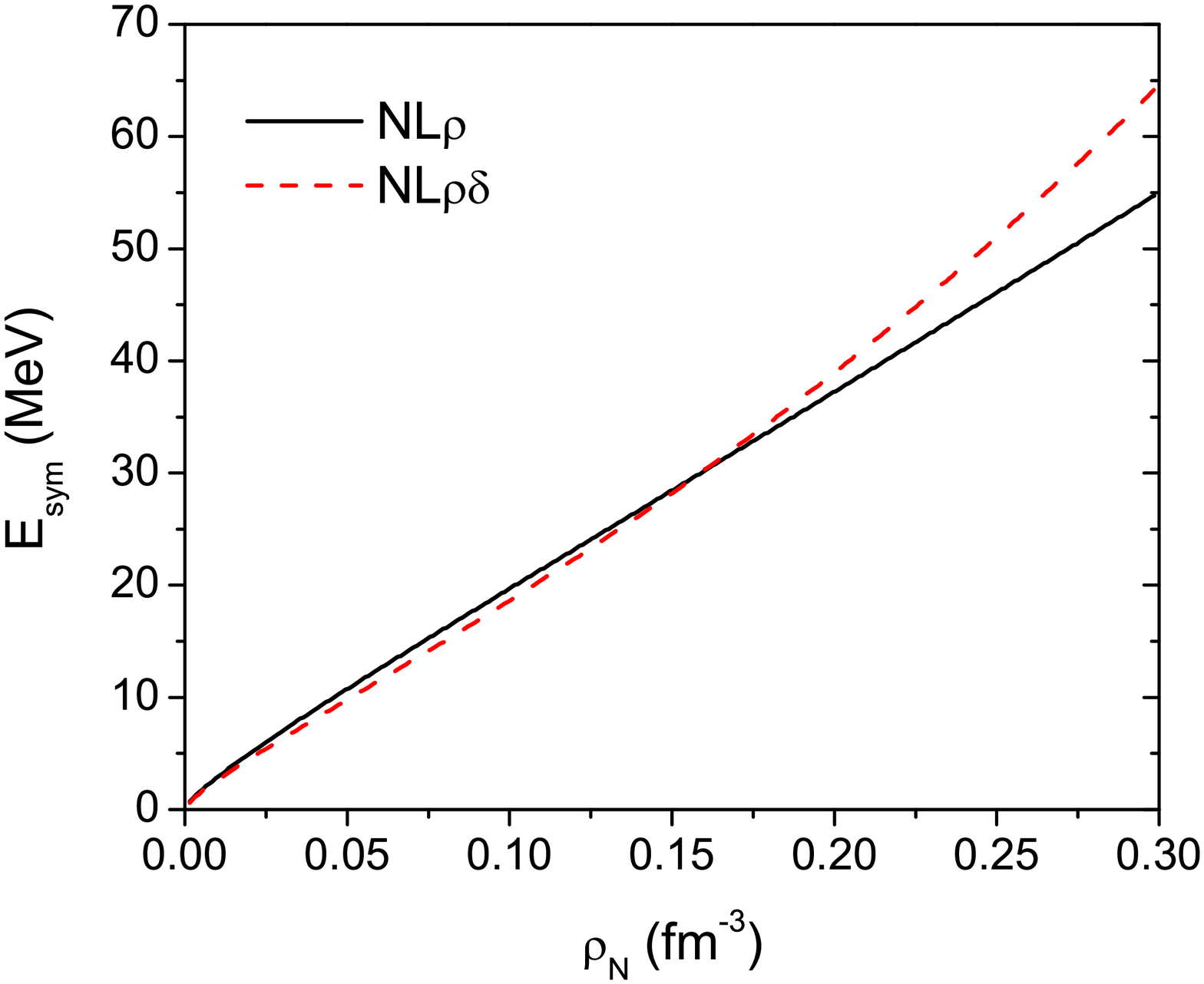}}
\caption{(Color online) Symmetry energy as a function of nuclear density in the $NL\rho$ and $NL\rho\delta$ models~\cite{Liu:2001iz}.}
\label{symE}
\end{figure}

Figure \ref{symE} shows the symmetry energy as a function of nuclear density obtained in the $NL\rho$ and $NL\rho\delta$ models. Compared to that in the $NL\rho$ model, the symmetry energy in the $NL\rho\delta$ model is slightly smaller below the saturation density and larger above it. Both are, however, about 30~MeV at saturation density.

\section{Relativistic Vlasov-Uehling-Uhlenbeck equation}\label{rvuu}

The time evolution of the nucleon distribution function $f(\vec{r},\vec{p};t)$ in phase space under the influence of the relativistic mean fields of the provious section is described by the relativistic Vlasov-Uehling-Uhlenbeck (RVUU) equation, which can be concisely written as~\cite{Ko:1996yy}
\begin{eqnarray}\label{transport}
\frac{\partial}{\partial t}f+\vec{v}\cdot\nabla_r f-\nabla_r H\cdot\nabla_p f=\mathcal{C}.
\end{eqnarray}
In the above, $H=\sqrt{M^{*2}+p^{*2}}+g_\omega\omega^0\pm g_\rho(\rho_3)_0$ is the Hamiltonian of a nucleon in the presence of mean fields with the upper sign for proton and the lower sign for neutron, and $\mathcal{C}$ denotes the collision integral. For the latter, we include both nucleon-nucleon elastic and inelastic scattering that produces a $\Delta$ resonance.

For the $\Delta$ resonance, which has an isospin of three halves, we assume that its interactions with the $\delta$ and $\rho$ fields are related to those of nucleons via its isospin structure in terms of those of the nucleon and the pion. For example, a $\Delta^+$ is related to the nucleon and pion by
\begin{eqnarray}
\bigg|\frac{3}{2}\frac{1}{2};\Delta^+\bigg\rangle&=&\sqrt{\frac{2}{3}}~\bigg|\frac{1}{2}\frac{1}{2};p\bigg\rangle\bigg|10;\pi^0\bigg\rangle\nonumber\\
&&+\sqrt{\frac{1}{3}}~\bigg|\frac{1}{2}\frac{-1}{2};n\bigg\rangle\bigg|11;\pi^+\bigg\rangle,
\end{eqnarray}
and its coupling to $\delta$ and $\rho$ has contributions of two thirds from proton and one third from neutron.
The effective mass and canonical energy-momentum of $\Delta^+$ are thus
\begin{eqnarray}
m^*_{\Delta^{+}}&=&m_{\Delta} -g_\sigma\sigma -\frac{1}{3}g_\delta\delta_3,\nonumber\\
p^\mu_{\Delta^+}&=&p^{\mu *} +g_\omega\omega^\mu+\frac{1}{3} g_\rho\rho_3^\mu.
\end{eqnarray}
Similarly, the effective mass and canonical energy-momentum of other charged states of $\Delta$ resonance are
\begin{eqnarray}
&&m^*_{\Delta^{++}}=m_{\Delta} -g_\sigma\sigma -g_\delta\delta_3,\nonumber\\
&&m^*_{\Delta^{0}}=m_{\Delta} -g_\sigma\sigma +\frac{1}{3}g_\delta\delta_3,\nonumber\\
&&m^*_{\Delta^{-}}=m_{\Delta} -g_\sigma\sigma +g_\delta\delta_3,
\label{deltam}
\end{eqnarray}
and
\begin{eqnarray}
p^\mu_{\Delta^{++}}&=&p^{\mu *} +g_\omega\omega^\mu+ g_\rho\rho_3^\mu,\nonumber\\
p^\mu_{\Delta^0}&=&p^{\mu *} +g_\omega\omega^\mu-\frac{1}{3} g_\rho\rho_3^\mu,\nonumber\\
p^\mu_{\Delta^-}&=&p^{\mu *} +g_\omega\omega^\mu- g_\rho\rho_3^\mu.
\label{deltap}
\end{eqnarray}
The time evolution of $\Delta$ resonance phase space distribution functions is then described by a similar RVUU equation as the one for nucleons.

For the collision integral in the RVUU equation, which changes the momenta of nucleons and $\Delta$ resonances, we use the baryon-baryon total elastic scattering
\begin{eqnarray}
\sigma_{BB\rightarrow BB}^{\rm elastic}({\rm mb})=55,~~~~~~~~~~~~~~\sqrt{s}<1.8993~{\rm GeV},\nonumber\\
=20+\frac{35}{1+100(\sqrt{s}-1.8993)},~\sqrt{s}\geq 1.8993~{\rm GeV}
\end{eqnarray}
and differential cross section
\begin{eqnarray}
\frac{d\sigma_{BB\rightarrow BB}^{\rm elastic}}{dt}\sim \exp\bigg[\frac{6\{3.65(\sqrt{s}-1.866)\}^6}{1+\{3.65(\sqrt{s}-1.866)\}^6}~t\bigg]
\end{eqnarray}
parameterized in Ref.~\cite{Bertsch:1988ik}.

For $\Delta$ resonance production in nucleon-nucleon scattering, we use the cross section calculated in the one-boson exchange model~\cite{Huber:1994ee} and employ the detailed balance to obtain its absorption cross section by nucleon~\cite{Danielewicz:1991dh,Li:1995pra}. In particular, we include only the reaction $NN\leftrightarrow N\Delta$ as heavy ion collisions energies considered in the present study are relatively low. Because of its resonance nature, a $\Delta$ is produced with a mass distribution of
\begin{eqnarray}
f_\Delta(m)=\frac{\Gamma^2(q)/4}{(m-m_0)^2+\Gamma^2(q)/4},
\end{eqnarray}
with $m_0=1.231~{\rm GeV}$ and a width given by
\begin{eqnarray}
\Gamma(q)=\frac{0.47}{1+0.6(q/m_\pi)^2}\frac{q^3}{m_\pi^2},
\label{width}
\end{eqnarray}
where $m_\pi$ and $q$ are, respectively, the pion mass and its three momentum from the decay $\Delta\rightarrow N\pi$ in the $\Delta$ rest frame.

For pions from decays of $\Delta$ resonances, they are treated in the RVUU model as free propagating particles except inelastic scattering with nucleons to form $\Delta$ resonances. The cross section for $\Delta$ formation in pion-nucleon scattering has the Breit-Wigner form,
\begin{eqnarray}
\sigma_{\pi N\rightarrow \Delta}=\sigma_{\rm max}\bigg(\frac{q_0}{q}\bigg)^2\frac{\Gamma^2(q)/4}{(\sqrt{s}-m_0)^2+\Gamma^2(q)/4},
\end{eqnarray}
where the maximum cross section $\sigma_{\rm max}$ is taken to be 190, 50, and 30 mb for $\pi^+p\rightarrow \Delta^{++}$ and $\pi^-n\rightarrow \Delta^{-}$, for $\pi^0 p\rightarrow \Delta^{+}$ and $\pi^0 n\rightarrow \Delta^{0}$, and for $\pi^-p\rightarrow \Delta^{0}$ and $\pi^+n\rightarrow \Delta^{+}$, respectively; $q$ is same as in Eq.~(\ref{width}) and $q_0$ is the three momentum of pion at $\sqrt{s}=m_0$~\cite{Li:1995pra}.

The phase space distribution function of pions thus satisfies a similar equation as given by Eq.(\ref{transport}) but without the mean-field term. There have been attempts to include the pion mean-field potential on pion production in heavy ion collisions~\cite{Xiong:1993pd}. Although the effect of pion mean-field potential on the charged pion ratio is not negligible~\cite{Xu:2009fj,Xu:2013aza}, it is nontrivial to be included in the transport model and will thus not be addressed in the present study.

We note that in both $NN\leftrightarrow N\Delta$ and $\Delta\leftrightarrow N\pi$ processes, the Pauli blocking for final-state baryons
is taken into account via the method of Ref.~\cite{Bertsch:1988ik}. Specifically, we determine the numbers of baryons $N_3$ and $N_4$ in the phase space volume defined by $|\Delta \vec{r}~|=[3/(4\pi\rho_0)]^{1/3}$ and $|\Delta \vec{ p}~|=[6\pi^2\rho_0/(2s+1)]^{1/3}$, with $s$ being the spin of baryon, around the phase space points (${\bf r}_3,{\bf p}_3$) and (${\bf r}_4,{\bf p}_4$) of the two final baryons and then take the probability of not being Pauli blocked to be $(1-N_3)(1-N_4)$.

\section{covariant threshold effect}\label{covariant}

For inelastic reactions in a medium such as $NN\to N\Delta$, their thresholds can be different from those in free space when mean-field potentials in the initial and final states are different. As shown in Appendix~\ref{thresholdE}, the threshold energy for the reaction $NN\to N\Delta$ in a medium described by the relativistic mean fields is determined by requiring the kinetic momenta of final nucleon and $\Delta$ in the frame where their total  mechanical momentum vanishes, i.e., $\bf{p}_{3}^{~*}+\bf{p}_{4}^{~*}=0$~\cite{Ferrini:2005jw}, to be zero, that is
\begin{eqnarray}
\sqrt{s}_{\rm th}=\sqrt{(m_3^*+\Sigma_3^0+m_4^*+\Sigma_4^0)^2-|\boldsymbol\Sigma_3+\boldsymbol\Sigma_4|^2}.
\label{threshold2}
\end{eqnarray}
For a static nuclear matter, i.e., $\boldsymbol\Sigma_i= 0$ and $\bf{p}_i^*\simeq 0$, the difference between incident and threshold energies~\cite{Ferrini:2005jw} is
\begin{eqnarray}
\sqrt{s}_{\rm in}-\sqrt{s}_{\rm th}\simeq E_1^*+E_2^*+\Sigma_1^0+\Sigma_2^0\nonumber\\
-m_3^*-m_4^*-\Sigma_3^0-\Sigma_4^0,
\end{eqnarray}
which in nonrelativistic limit becomes
\begin{eqnarray}
\sqrt{s}_{\rm in}-\sqrt{s}_{\rm th}\simeq m_1+m_2-m_3-m_4\nonumber\\
+\Sigma_1^s+\Sigma_2^s-\Sigma_3^s-\Sigma_4^s+\frac{|\bf{p}_1^*|^2}{2m_1^*}+\frac{|\bf{p}_2^*|^2}{2m_2^*}\nonumber\\
+\Sigma_1^0+\Sigma_2^0-\Sigma_3^0-\Sigma_4^0
\label{sdiff}
\end{eqnarray}
with $\Sigma_i^s=m_i^*-m_i$.

\begin{table}[h]
\centering
\begin{tabular}{c| c | c}
\hline
scattering & ~$\Sigma_1^s+\Sigma_2^s-\Sigma_3^s-\Sigma_4^s$~ & ~$\Sigma_1^\mu+\Sigma_2^\mu-\Sigma_3^\mu-\Sigma_4^\mu$~ \\[2pt]
\hline
elastic & \\[2pt]
~$NN\rightarrow NN$~& ~0~ & ~0~\\[2pt]
~$N\Delta\rightarrow N\Delta$~& ~0~ & ~0~\\[2pt]
~$\Delta\Delta\rightarrow \Delta\Delta$~& ~0~ & ~0~\\[2pt]
inelastic & \\[2pt]
~$pp\rightarrow n\Delta^{++}$~& ~$-2g_\delta\delta_3$~ & ~$2g_\rho\rho_3^\mu$~ \\[2pt]
~$pp\rightarrow p\Delta^{+}$~& ~$-(2/3)g_\delta\delta_3$~ & ~$(2/3)g_\rho\rho_3^\mu$~ \\[2pt]
~$pn\rightarrow n\Delta^{+}$~& ~$-(2/3)g_\delta\delta_3$~ & ~$(2/3)g_\rho\rho_3^\mu$~ \\[2pt]
~$pn\rightarrow p\Delta^{0}$~& ~$(2/3)g_\delta\delta_3$~ & ~$-(2/3)g_\rho\rho_3^\mu$~ \\[2pt]
~$nn\rightarrow n\Delta^{0}$~& ~$(2/3)g_\delta\delta_3$~ & ~$-(2/3)g_\rho\rho_3^\mu$~ \\[2pt]
~$nn\rightarrow p\Delta^{-}$~& ~$2g_\delta\delta_3$~ & ~$-2g_\rho\rho_3^\mu$~ \\[2pt]
\hline
decay & ~$\Sigma_1^s-\Sigma_2^s$~ & ~$\Sigma_1^\mu-\Sigma_2^\mu$~\\[2pt]
\hline
~$\Delta^{++}\rightarrow p\pi^{+}$~& ~0~& ~0~ \\[2pt]
~$\Delta^+\rightarrow p\pi^{0}$~& ~$(2/3)g_\delta\delta_3$~& ~$-(2/3)g_\rho\rho_3^\mu$~ \\[2pt]
~$\Delta^+\rightarrow n\pi^{+}$~& ~$-(4/3)g_\delta\delta_3$~& ~$(4/3)g_\rho\rho_3^\mu$~ \\[2pt]
~$\Delta^0\rightarrow p\pi^{-}$~& ~$(4/3)g_\delta\delta_3$~& ~$-(4/3)g_\rho\rho_3^\mu$~ \\[2pt]
~$\Delta^0\rightarrow n\pi^{0}$~& ~$-(2/3)g_\delta\delta_3$~& ~$(2/3)g_\rho\rho_3^\mu$~ \\[2pt]
~$\Delta^-\rightarrow n\pi^{-}$~& ~0~& ~0~ \\[2pt]
\hline
\end{tabular}
\caption{Differenece between initial and final scalar and vector mean fields in nucleon-nucleon elastic and inelastic scatterings as well as in decays of $\Delta$ resonances.}
\label{threshold}
\end{table}

Table \ref{threshold} shows the differences between initial and final scalar and vector mean fields in nucleon-nucleon elastic and inelastic scatterings as well as in decays of $\Delta$ resonances. A positive difference reduces the threshold for a reaction, while a negative difference increases its threshold. Since $\delta_3$ and $\rho_3^0$ are negative in neutron-rich matter, the change of the threshold due to the scalar mean field thus enhances the production of $\Delta^+$ and $\Delta^{++}$ and suppresses that of $\Delta^0$ and $\Delta^-$, while the effect of the vector mean field is opposite.

In the $NL\rho$ model, where $g_\delta=0$, only $\Sigma^\mu$ can be different between initial and final states. As a result, the production of $\Delta^0$ and $\Delta^-$ is enhanced and that of $\Delta^+$ and $\Delta^{++}$ is suppressed. Because the former mainly decay to $N\pi^-$ and the latter mainly to $N\pi^+$, the $\pi^-/\pi^+$ ratio thus increases in heavy ion collisions. Also, the difference in $\Sigma^\mu$ between the initial and finite states favors $\Delta^+\rightarrow p\pi^{0}$ over $\Delta^+\rightarrow n\pi^{+}$ and $\Delta^0\rightarrow p\pi^{-}$ over $\Delta^0\rightarrow n\pi^{0}$.

In the $NL\rho\delta$ model, both $\Sigma^s$ and $\Sigma^\mu$ can be different between initial and final states. However, the difference in $\Sigma^\mu$ is larger than that in $\Sigma^s$. For example, in nuclear matter of $\rho_N=0.3~{\rm fm^{-3}}$ and $\alpha=0.5$, $\Sigma^0$ differs by 373 MeV and $\Sigma^s$ by $-122$ MeV between the initial and final states. The net difference of 251 MeV is larger than the 112 MeV in the difference of $\Sigma^0$ in the $NL\rho$ model. Therefore, the $\pi^-/\pi^+$ ratio is expected to be more enhanced in the $NL\rho\delta$ than in the $NL\rho$ model.

\section{results}\label{results}

Including the threshold effects on $\Delta$ resonance production based on the nonlinear relativistic mean-field models, we study in this Section Au+Au collisions at energies from 0.3 AGeV to 0.7 AGeV and compare results on the charged pion ratio with the experimental data from the FOPI Collaboration. The RVUU equations are solved using the test particle method~\cite{Bertsch:1988ik} with initial conditions that nucleons inside in each nucleus are distributed according to the Wood-Saxon form,
\begin{eqnarray}
\rho(\vec{r})\sim \frac{1}{1+\exp[(r-c)/a]},
\label{Wood-Saxon}
\end{eqnarray}
with the parameters $c=6.38~{\rm fm}$ and $a=0.535~{\rm fm}$ for $^{197}{\rm Au}$ nucleus~\cite{De Jager:1987qc}, and their momentum-space distributions are obtained by filling up to the local Fermi momentum.

\begin{figure}[h]
\centerline{
\includegraphics[width=9 cm]{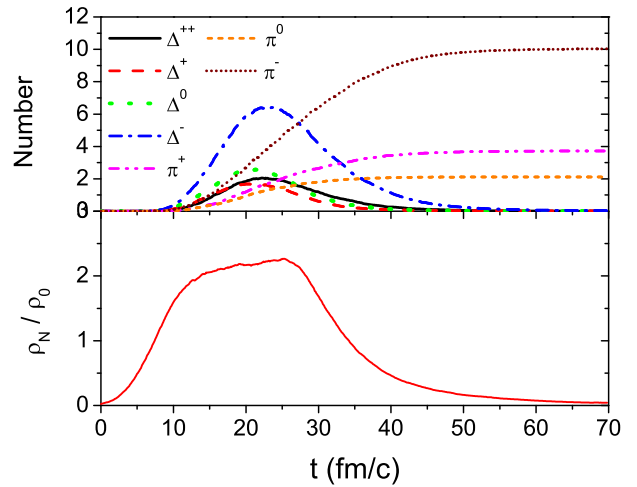}}
\caption{(Color online) Numbers of $\Delta$ baryons and pions (upper panel) and central density divided by the saturation density (lower panel) in the $NL\rho$ model as functions of time in Au+Au collisions at impact parameter of $1~{\rm fm}$ and energy of $E/A=400~{\rm MeV}$.}
\label{profile1}
\end{figure}

Figure~\ref{profile1} shows the numbers of $\Delta$ baryons and pions as functions of time for Au+Au collisions at the impact parameter of $1~{\rm fm}$ and energy of $E/A=400~{\rm MeV}$ in the $NL\rho$ model. It is seen that most $\Delta$ baryons are produced during the high density stage as shown in the lower panel of the figure.

\begin{figure}[h]
\centerline{
\includegraphics[width=9 cm]{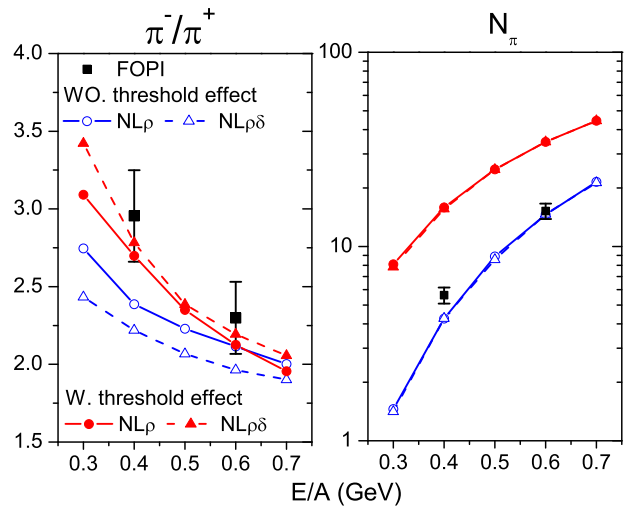}}
\caption{(Color online) $\pi^-/\pi^+$ ratio (left panel) and pion yields (right panel) as functions of collision energy with and without the threshold effect in Au+Au collisions at impact parameter of $1~{\rm fm}$ from the $NL\rho$ and $NL\rho\delta$ models. Experimental data are from the FOPI Collaboration~\cite{Reisdorf:2006ie}.}
\label{fopi1}
\end{figure}

Figure~\ref{fopi1} shows the $\pi^-/\pi^+$ ratio and the total pion yield as functions of collision energy in Au+Au collisions at impact parameter of $1~{\rm fm}
$ from the $NL\rho$ and $NL\rho\delta$ models. For the case of without threshold effect, nucleon and $\Delta$ scatterings and $\Delta$ decays are treated
as if there are no mean-field potentials, that is taking $m_i^*=m_i$ and $p_i^{\mu~*}=p_i^\mu$. In this case, the $\pi^-/\pi^+$ ratio is larger for the $NL\rho$
model than for the $NL\rho\delta$. This is because the high density matter formed in these collisions, where most $\Delta$ baryons are produced, is more
neutron-rich when the $NL\rho$ model, which has a softer symmetry energy than the $NL\rho\delta$ model, is used. Since there are more neutron-neutron scattering in more neutron-rich matter, more $\Delta^-$ and thus $\pi^-$ are produced. As a result, the $\pi^-/\pi^+$ ratio is larger for the $NL\rho$ model as shown by the lower two curves in the left panel of Fig.~\ref{fopi1}, similar to that found in Ref.~\cite{Gaitanos:2003zg}. The symmetry energy effect on the $\pi^-/\pi^+$ ratio is, however, reversed by the threshold effect as shown by the upper two curves in the left panel of Fig.~\ref{fopi1}. This is due to the fact discussed in the previous section that the threshold effect enhances $\pi^-$ production and suppresses $\pi^+$ production in neutron-rich nuclear matter in both $NL\rho$ and $NL\rho\delta$ models, and the effect is larger for the $NL\rho\delta$ model than for the $NL\rho$ model. The $\pi^-/\pi^+$ ratio thus increases for both models but more for the $NL\rho\delta$ model than for the $NL\rho$ model~\cite{Ferrini:2005jw}. Figure~\ref{fopi1} further shows that including the threshold effect helps reproduce the experimental data on the $\pi^-/\pi^+$ ratio, although the effect becomes smaller as the collision energy increases.

Because of increased nucleon kinetic energy due to its reduced mass in nuclear matter, the threshold effect increases the total pion yield compared to that without including the threshold effect as shown in the left panel of Fig.~\ref{fopi1}. Since the total pion yield in the case without the threshold effect is close to the experimental data, including the threshold effect leads to an overestimate of the total pion yield. To reproduce the experimental data, we take into account medium effects on the cross section for $\Delta$ production by assuming the following density dependence:
\begin{eqnarray}
\sigma_{NN\rightarrow \Delta N}(\rho_N)=\sigma_{NN\rightarrow \Delta N}(0)\exp(-A\rho_N/\rho_0),
\end{eqnarray}
where $\rho_N$ is the nucleon density and $A$ is a fitting parameter. The cross section for the inverse reaction also becomes density dependent and is determined by the detailed balance. The density-dependent factor suppresses pion production in nuclear matter, and it is consistent with those found in other studies~\cite{TerHaar:1987ce,Larionov:2003av,Prassa:2007zw}.

\begin{figure}[h]
\centerline{
\includegraphics[width=9 cm]{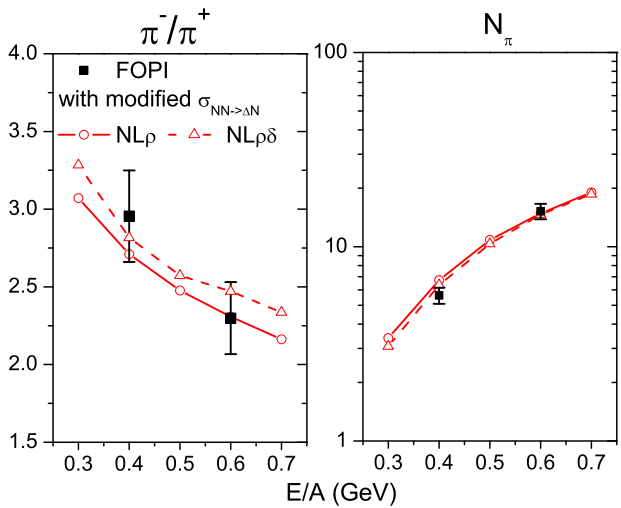}}
\caption{(Color online) $\pi^-/\pi^+$ ratio (left panel) and total pion yield (right panel) as functions of collision energy obtained with the threshold effect as well as density-dependent $\Delta$ production cross in Au+Au collisions at impact parameter of $1~{\rm fm}$ for both the $NL\rho$ and $NL\rho\delta$ models. Experimental data are from the FOPI Collaboration~\cite{Reisdorf:2006ie}.}
\label{fopi2}
\end{figure}

We find that the experimentally measured total pion yield can be described very well by taking $A=1.2$ in the density-dependent $\Delta$ production cross section for both the $NL\rho$ and $NL\rho\delta$ models as shown in the right panel of Fig.~\ref{fopi2}. The resulting $\pi^-/\pi^+$ ratio is shown in the left panel of Fig.~\ref{fopi2}, and it is seen to differ only slightly from those obtained with the vacuum $\Delta$ production cross section.

\section{summary}\label{summary}

Using the relativistic Vlasov--Uehling-Uhlenbeck equation based on the nonlinear relativistic mean-field models, we have studied the covariant threshold effect on the pion yield and the $\pi^-/\pi^+$ ratio in Au+Au collisions. We have found that besides enhancing the pion yield and the $\pi^-/\pi^+$ ratio, the threshold effect also reverses the effect of nuclear symmetry energy on the $\pi^-/\pi^+$ ratio. Although including the threshold effect leads to a better description of the measured $\pi^-/\pi^+$ ratio from the FOPI Collaboration, it gives too large a total pion yield compared to the experimental data. Introducing a density-dependence in the $\Delta$ production cross sections, we have been able to describe both the pion yield and the $\pi^-/\pi$ ratio measured in experiments. The large errors in the experimentally measured $\pi^-/\pi^+$ ratio prevent, however, the distinction between the predictions from the $NL\rho$ and $NL\rho\delta$ models, which correspond to a soft and a stiff nuclear symmetry energy, respectively. Since the in-medium threshold effect has an opposite effect on the $\pi^-/\pi^+$ ratio in heavy ion collisions from the effect due to the stiffness of nuclear symmetry energy at high density, it is important to include this effect in extracting the high-density behavior of nuclear symmetry energy from experimentally measured $\pi^-/\pi^+$ ratio.

\section*{Acknowledgements}

We thank Massimo Di Toro for helpful communications. This work was supported in part by the U.S. National Science Foundation under Grant No. PHY-1068572 and the Welch Foundation under Grant No. A-1358 as well as by the DFG.



\appendix
\section{symmetry energy}\label{symmetryE}

Neglecting $\sigma$ and $\omega$ fields, which depend only on the total nuclear density $\rho_N$ and thus are not relevant to the symmetry energy, the energy density of nuclear matter at zero temperature is
\begin{eqnarray}
\epsilon(\rho_N,\rho_I)=\sum_{i=p,n}2\int^{p_F^i}\frac{d^3p}{(2\pi)^3}\sqrt{m_i^{*2}+p^2}\nonumber\\
+\frac{1}{8f_\delta}(m_n^*-m_p^*)^2+\frac{1}{2}f_\rho\rho_I^2+\cdots,
\end{eqnarray}
where $\delta_3=(m_n^*-m_p^*)/(2g_\delta)$ and $(\rho_{3})_0=(g_\rho/m_\rho^2)\rho_I$.

The derivative of $\epsilon$ with respect to $\rho_I$ is then
\begin{eqnarray}
\frac{\partial \epsilon}{\partial \rho_I}=\frac{(p_F^p)^2E_F^p}{\pi^2}\frac{\partial p_F^p}{\partial \rho_I}+\frac{(p_F^n)^2E_F^n}{\pi^2}\frac{\partial p_F^n}{\partial \rho_I}\nonumber\\
+\phi_p\frac{\partial m_p^*}{\partial \rho_I}+\phi_n\frac{\partial m_n^*}{\partial \rho_I}+f_\rho\rho_I\nonumber\\
+\frac{1}{4f_\delta}(m_n^*-m_p^*)\bigg(\frac{\partial m_n^*}{\partial \rho_I}-\frac{\partial m_p^*}{\partial \rho_I}\bigg),
\label{derivative1}
\end{eqnarray}
where $p_F^i$ and $E_F^i$ are the Fermi momentum and energy of nucleon type $i$, respectively, and $\phi_i$ is the scalar density defined in Eq.~(\ref{source}), which has the explicit expression
\begin{eqnarray}
\phi_i=\frac{m_i^*}{2\pi^2}\bigg[E_F^ip_F^i-m_i^{*2}\ln\bigg(\frac{E_F^i+p_F^i}{m_i^*}\bigg)\bigg]
\label{condensate}
\end{eqnarray}
at zero temperature.

In terms of the relations
\begin{eqnarray}
j_p^0=\frac{1}{2}(\rho_N+\rho_I)=\frac{1}{3\pi^2}(p_F^p)^3,\nonumber\\
j_n^0=\frac{1}{2}(\rho_N-\rho_I)=\frac{1}{3\pi^2}(p_F^n)^3,
\end{eqnarray}
the derivatives of the Fermi momentum of protons and neutrons with respect to the isospin density $\rho_I$ are,
respectively,
\begin{eqnarray}
\frac{\partial p_F^p}{\partial \rho_I}=\frac{\pi^2}{2(p_F^p)^2},~~~~~~~~\frac{\partial p_F^n}{\partial \rho_I}=-\frac{\pi^2}{2(p_F^n)^2}.
\end{eqnarray}
Using Eq.~(\ref{momentum})$-$(\ref{meson2}), and (\ref{condensate}), the derivatives of the proton and neutron masses with the isospin density are then
\begin{eqnarray}
\bigg(1+f_\delta A_p\bigg)\frac{\partial m_p^*}{\partial \rho_I}=-\frac{f_\delta m_p^*}{2E_F^p}-\frac{f_\delta m_n^*}{2E_F^n}+f_\delta A_n\frac{\partial m_n^*}{\partial \rho_I},\nonumber\\
\bigg(1+f_\delta A_n\bigg)\frac{\partial m_n^*}{\partial \rho_I}=\frac{f_\delta m_p^*}{2E_F^p}+\frac{f_\delta m_n^*}{2E_F^n}+f_\delta A_p\frac{\partial m_p^*}{\partial \rho_I},\nonumber\\
\label{simultaneous}
\end{eqnarray}
where
\begin{eqnarray}
A_p(p_F^p,m_p^*)=\frac{\partial \phi_p}{\partial m_p^*},~~~~~~A_n(p_F^n,m_n^*)=\frac{\partial \phi_n}{\partial m_n^*}.\nonumber
\end{eqnarray}
In symmetric nuclear matter ($\rho_I=0$), $A_p=A_n$ and $A_p+A_n=A(p_F,m^*)$ reduces to Eq.~(\ref{aa}).

Solving the simultaneous equations (\ref{simultaneous}) gives
\begin{eqnarray}
\frac{\partial m_p^*}{\partial \rho_I}=-\frac{f_\delta}{1+f_\delta(A_p+A_n)}\bigg(\frac{m_p^*}{2E_F^p}+\frac{m_n^*}{2E_F^n}\bigg),\nonumber\\
\frac{\partial m_n^*}{\partial \rho_I}=-\frac{\partial m_p^*}{\partial \rho_I}.~~~~~~~~~~~~~~~~~~~~~~~~~~~~~~~~~~~~~
\end{eqnarray}
Therefore, Eq.~(\ref{derivative1}) vanishes at $\rho_I=0$, that is
\begin{eqnarray}
\frac{\partial \epsilon}{\partial \rho_I}\bigg|_{\rho_I=0}=0.
\end{eqnarray}

The nuclear symmetry energy is obtained in the same way by taking the second derivative of $\epsilon$ with respect to $\rho_I$, and the result is
\begin{eqnarray}
E_{\rm sym}(\rho_N)=\frac{1}{2}\rho_N\frac{\partial^2 \epsilon}{\partial \rho_I^2}\bigg|_{\rho_I=0}~~~~~~~~~~~~~~~~~~~~~~~~~\nonumber\\
=\frac{p_F^2}{6E_F}+\frac{\rho_N}{2}\bigg[f_\rho-\frac{f_\delta m^{*2}}{E_F^2\{1+f_\delta A(p_F, m^*)\}}\bigg],
\end{eqnarray}
where $p_F^p=p_F^n=p_F$ and $m_p^*=m_n^*=m^*$.

\section{threshold energy in a medium}\label{thresholdE}

To illustrate how mean fields affect the threshold energy of a reaction in a medium, we consider for simplicity the decay of $\Delta^-$ into $n$ and $\pi^-$. In the frame of ${\bf p}_{\Delta-}=0$, the energy conservation gives
\begin{eqnarray}
\sqrt{s}=\sqrt{m_{\Delta-}^{*2}+|\boldsymbol\Sigma_{\Delta-}|^2}+\Sigma_{\Delta-}^0~~~~~~~~~~~~~~~~~~~\nonumber\\
=\sqrt{m_n^{*2}+|{\bf p}-\boldsymbol\Sigma}_n|^2+\Sigma_n^0+\sqrt{m_\pi^2+|{\bf p}|^2},
\label{s1}
\end{eqnarray}
assuming the pion mass and momentum are not affected by nuclear mean fields. Requiring the threshold for a $\Delta^-$ to decay to $n$ and $\pi^-$ corresponds to $|{\bf p}|=0$ as in free space, the minimum $\Delta^-$ mass is then
\begin{eqnarray}
m_{\Delta-}^{*2}=m_n^{*2}+m_\pi^{2}+2m_\pi\sqrt{m_n^{*2}+|\vec{\Sigma}_n|^2},
\label{threshold1}
\end{eqnarray}
where $\Sigma_{\Delta-}^\mu=\Sigma_n^\mu$ from Eq.~(\ref{deltap}).

On the other hand, the minimum $\Delta^-$ mass can be found by minimizing Eq.~(\ref{s1}) with respect to $|{\bf p}|$, that is
\begin{eqnarray}
\frac{\partial\sqrt{s}}{\partial|{\bf p}|}=0,
\end{eqnarray}
which leads to
\begin{eqnarray}
\bigg\{m_n^{*2}-m_\pi^2+|\boldsymbol\Sigma_n|^2\sin^2\theta\bigg\}|{\bf p}|^2~~~~~~~~~~~~~\nonumber\\
+2m_\pi^2|\boldsymbol\Sigma_n|\cos\theta|{\bf p}|
-m_\pi^2|\boldsymbol\Sigma_n|^2\cos^2\theta=0,
\end{eqnarray}
where $\theta$ is the angle between ${\bf p}$ and $\boldsymbol\Sigma_n$.
If $\cos\theta=0$, i.e., ${\bf p}$ and $\boldsymbol\Sigma_n$ are perpendicular, then $
{\bf p}=0$ and the minimum value for $\Delta^-$ mass is that given by Eq.(\ref{threshold1}). For $\cos\theta=1$, i.e., $
{\bf p}$ and $\boldsymbol\Sigma_n$ are in the same direction, $\sqrt{s}$ has instead the
minimum value
\begin{eqnarray}
\sqrt{s}=\sqrt{(m_n^*+m_\pi)^2+|\boldsymbol\Sigma_n|^2}+\Sigma_n^0,
\end{eqnarray}
at $|{\bf p}|=m_\pi\boldsymbol\Sigma_n/(m_n^*+m_\pi)$ or
\begin{eqnarray}
m_{\Delta-}^{*}=m_n^{*}+m_\pi,
\end{eqnarray}
according to Eq. (\ref{s1}), which is less than that given in Eq. (\ref{threshold1}). Therefore, the minimum or threshold $\Delta^-$ mass does not take place at $|{\bf p}|=0$ but at finite ${\bf p}$ that is in the same direction as $\boldsymbol\Sigma_n$ in the frame of ${\bf p}_{\Delta-}=0$, which is equivalent to $|{\bf p}^*|=0$ in the frame of ${\bf p}_{\Delta-}^{~*}=0$.

\end{document}